\documentclass[11pt]{article}
\usepackage{amsmath,amssymb,color,graphics,epsfig,cite}

\usepackage{amsfonts}

\newcommand{\be}{\begin{equation}}
\newcommand{\ee}{\end{equation}}
\newcommand{\bea}{\setlength\arraycolsep{2pt} \begin{eqnarray}}
\newcommand{\eea}{\end{eqnarray}}
\newcommand{\nn}{\nonumber}

\def\ft#1#2{{\textstyle{\frac{\scriptstyle #1}{\scriptstyle #2} } }}
\def\fft#1#2{{\frac{#1}{#2}}}

\def\0{{\sst{(0)}}}
\def\1{{\sst{(1)}}}
\def\2{{\sst{(2)}}}
\def\3{{\sst{(3)}}}
\def\4{{\sst{(4)}}}
\def\5{{\sst{(5)}}}
\def\6{{\sst{(6)}}}
\def\7{{\sst{(7)}}}
\def\8{{\sst{(8)}}}
\def\sst#1{{\scriptscriptstyle #1}}

\begin{document}

\begin{center}
{\Large {\bf Schwarzschild Black Hole in Uniform Magnetic Field\\ in Einstein-Maxwell-Dilaton Theories}}

\vspace{20pt}

Liang Ma, Peng-Yu Wu, H.~L\"u

\vspace{10pt}

{\it Center for Joint Quantum Studies, Department of Physics,\\
School of Science, Tianjin University, Tianjin 300350, China }

\vspace{40pt}

\underline{ABSTRACT}
\end{center}

We obtain the Kerr black hole in an external uniform magnetic field in the Kaluza-Klein (KK) theory via a standard solution-generating technique. Using this as our starting point, we construct exact static solutions describing a Schwarzschild black hole immersed in the external uniform magnetic field in general string-inspired Einstein-Maxwell-dilaton (EMD) theories, thereby embedding the static Kerr-Bertotti-Robinson (KBR) solution into the general framework of EMD theories. We also construct electrovacuum solutions involving both electric and magnetic fields in EMD theories and generalize the system to include mutltiple Maxwell fields and dilatons. These equations of motion can be reduced to Toda-like equations, allowing us to obtain analytic electrovacua associated with Toda equations of all rank-2 Lie groups. We further construct exact solutions of a Schwarzschild black hole immersed in these electrovacua.

\vfill {\footnotesize maliang0@tju.edu.cn\ \ \ wupy2023@tju.edu.cn\ \ \ mrhonglu@gmail.com}

%{\footnotesize \hoch{*}Corresponding author}

\thispagestyle{empty}
\pagebreak

\tableofcontents
\addtocontents{toc}{\protect\setcounter{tocdepth}{2}}

%\newpage

\section{Introduction}

Einstein-Maxwell gravity unites the two known long-range fundamental forces in nature, and a particular class of solutions consists of electrovacua, which describe vacuum spacetimes containing only external electric or magnetic fields, or both, but no intrinsic charges. The best known examples are Bonnor-Melvin (BM) \cite{Bonnor:1954tis,Melvin:1963qx} and Bertotti-Robinson (BR) \cite{Robinson:1959ev,Bertotti:1959pf} solutions. These solutions typically have cohomogeneity one or are even homogeneous, and are locally equivalent to solutions referred to as fluxbranes \cite{Gibbons:1987ps,Dowker:1993bt,Dowker:1994up, Dowker:1995gb, Russo:1995tj,Russo:1995ik,Emparan:2001gm,Chen:2001nr, Yazadjiev:2005gs, Ivashchuk:2013jja} in the string theory literature. For example, the BM solution is locally related to magnetic string solution, where an additional Killing direction along the string is present. (In the General Relativity literature, the string direction is usually referred to as the cylindrical coordinate.) The homogeneous BR electrovacuum is locally equivalent to the near-horizon geometry of an extremal Reissner-Nordstr\"om (RN) black hole. A new electrovacuum \cite{Ma:2026ima,Ma:2026uok} involving a cosmological horizon is locally related to a charged Kundt class \cite{Ovcharenko:2026uxi}, which is itself a certain special case and analytic continuation of the RN black hole.

However, constructing black holes immersed in an electrovacuum is very different from constructing non-extremal fluxbranes. In the former case, the resulting solution typically has higher cohomogeneity, and reduces to the usual Schwarzschild or Kerr black holes when the external electromagnetic fields are turned off. In the latter case, the non-extremality is achieved while preserving the same cohomogeneity, and removing the electromagnetic fields typically yields neutral black branes.  Nevertheless, the techniques developed for constructing fluxbrane are useful for constructing electrovacua and the corresponding black holes in theories beyond Einstein-Maxwell gravity.

The BR electrovacuum is special in that the electromagnetic field is constant. In other words, the Maxwell field-strength invariant is constant. Recently, exact solutions of Kerr black holes in the BR background (KBR) were constructed in \cite{Podolsky:2025tle, Ovcharenko:2025cpm,Ovcharenko:2026tos}, which inspired many subsequent works on the related subjects \cite{Astorino:2025lih,Astorino:2026okd,Ovcharenko:2026byw, Barrientos:2026shy,Hu:2026slp,Herdeiro:2026jem,DiPinto:2026rvp,Lyu:2026pqh,
Huang:2026qzj,Li:2026gvm,DiPinto:2026ynu,Astorino:2026nhd,Furugori:2026qbb}. Einstein-Maxwell gravity can naturally arise as a consistent truncation of supergravity theories, which are low energy effective theories of strings. This naturally raises the question of whether such solutions can be constructed in EMD gravity, which is a simple and natural generalization of Einstein-Maxwell gravity in string theory. The theory contains an additional dilaton, a natural component of string theory, which couples non-minimally to the Maxwell field through an exponential coupling. For the dilaton coupling constants $a=\sqrt3, 1,1/\sqrt3,0$, the EMD theory can arise from consistent truncations of supergravities. In string-inspired EMD theories, the dilaton coupling constant is typically allowed to take any real value. When $a=0$, the theory reduces to Einstein-Maxwell gravity coupled to a free massless scalar, from which the KBR solution is recovered. However, constructing the KBR-type solutions in a generic EMD theory may not be possible; in fact, Exact solutions of rotating black holes in general EMD theories remain unknown. (See \cite{Herdeiro:2025blx} for numerical solutions.) One goal of this paper is to construct a static configuration describing a Schwarzschild black hole immersed in an external uniform magnetic field, which reduces to the static KBR when $a=0$.

Our construction relies on two ingredients. One is to make use of the fact that the KK theory corresponds to an EMD theory with $a=\sqrt3$. This allows us to apply a standard solution-generating technique to obtain new solutions in the KK theory from a known Ricci-flat metrics in four dimensions. The standard Melvin-type black hole in the KK theory was indeed obtained by this procedure, using the Schwarzschild black hole as a seed solution \cite{Dowker:1995gb}. To achieve our goal, we also employ a second tool that is the recently constructed Ricci-flat metric, a $B$-deformed Kerr black hole, which can serve as a seed for constructing the KBR solution via a different solution-generating technique \cite{Ma:2026otg}. We find that the same Ricci-flat seed can give rise to the Kerr black hole immersed in an external uniform magnetic field in the KK theory. From the structure of the KK solution, we are able to construct directly exact Schwarzschild black hole solutions immersed in a uniform magnetic field in general EMD theories.

In four dimensions, the Maxwell field admits an electromagnetic duality transformation. In Einstein-Maxwell gravity, the equations of motion are invariant under the $SO(2)$ electromagnetic duality transformations. This symmetry breaks down to a $\mathbb Z_2$ symmetry, which makes the construction of electrovacua or fluxbranes involving both electric and magnetic fields in EMD theories highly nontrivial. We carry out this construction for general EMD theories. For some special dilaton couplings, we are able to construct the Schwarzschild black hole in such electrovacua.

The paper is organized as follows. In Section 2, we briefly review the four-dimensional EMD theory and its generalization to Einstein-Maxwell-Maxwell-Dilaton (EMMD) gravity. In Section 3, we construct a Kerr black hole immersed in an external uniform magnetic field in the KK theory. In Section 4, we construct a Schwarzschild black hole immersed in an external uniform magnetic field in general EMD theories. In Section 5, we construct electrovacua involving independent electric and magnetic fields in general EMD theories. We further obtain exact Schwarzschild black hole solutions in such electrovacua for  $a=1$ and $a=\sqrt3$. In Section 6, we consider the EMMD theory and construct electrovacua involving external electric fields of both Maxwell fields. We find that the equations of motion can be reduced to Toda-like equations and for suitable dilaton coupling constants, the equations become precisely Toda equations associated with all rank-2 Lie groups, which can be solved analytically. We further construct the Schwarzschild black hole in these special (analytic) electrovacua. We conclude the paper in Section 7.

\section{EMD gravity and its generalizations}
\label{sec:emdreview}

EMD theory describes Einstein gravity coupled to a Maxwell field $A_\1$ and a massless scalar $\phi$, with the scalar field non-minimally coupled to the Maxwell field through an exponential coupling. The Lagrangian is given by
\be
{\cal L}=\sqrt{-g} \Big(R - \ft12 (\partial\phi)^2 - \ft14 e^{a\phi} F_\2^2\Big)\,,\qquad F_\2=dA_\1\,,\label{emdlag}
\ee
where $a$ is the dilaton coupling constant. The scalar is called dilaton because the Lagrangian is invariant under a constant shifting of the scalar, accompanied by an appropriate rescaling of the Maxwell field, namely
\be
\phi \rightarrow \phi + c\,,\qquad A_\1\rightarrow e^{-\fft12 a c} A_\1\,.
\ee
Under the conformal transformation, $ds^2=e^{a\phi} d\tilde s^2$, the Lagrangian of the EMD theory becomes
\be
\widetilde {\cal L} = \sqrt{-\tilde g}\, e^{a\phi} \Big(\widetilde R + \ft32 a^2 (\partial\phi)^2 - \ft14 F_\2^2\Big)\,.\label{pframe}
\ee
We refer to $d\tilde s^2$ as the (electric) particle frame (PF) while the metric $ds^2$ associated with the Lagrangian \eqref{emdlag} is referred to as the Einstein frame (EF). It is easy to see that
\be
e^{2a \phi} F_\2^2\Big|_{\rm EF} = F_\2^2\Big|_{\rm PF}\,.\label{EFPF}
\ee
If an electromagnetic field is constant in the Einstein frame, i.e.~$F_\2^2\big|_{\rm EF}=$constant, we refer to it as a uniform electromagnetic field, whereas if it is constant in the particle frame, namely $F_\2^2\big|_{\rm PF}=$constant, we refer to it as a conformally-uniform electromagnetic field.

When $a=0$, the EMD theory reduces to Einstein-Maxwell gravity coupled to a free massless scalar, in which case, the theory is invariant under an $SO(2)$ electromagnetic duality at the level of equations of motion. For a generic non-vanishing dilaton coupling constant $a$, this one-parameter continuous symmetry reduces to a discrete $\mathbb Z_2$ electromagnetic duality, namely
\be
F_\2\rightarrow e^{-a\phi} {*F}_\2\,,\qquad \hbox{followed by}\qquad \phi\rightarrow -\phi\,.\label{emduality}
\ee
The absence of a continuous parameter in the electromagnetic duality implies that configurations involving both electric and magnetic fields are inequivalent to configurations with either a purely electric or purely magnetic field. One of the goals of this paper is to construct new electrovacua in the EMD theory. It follows from electromagnetic duality \eqref{emduality} and the Einstein-particle frame relation \eqref{EFPF} that the spacetime with uniform magnetic (electric) field in EMD theory is related by the electromagnetic duality to a spacetime with the conformally-uniform electric (magnetic) field.

As was observed in \cite{Lu:2013eoa}, the equations of motion of an EMD theory governing spacetime configurations involving both electric and magnetic fields are equivalent to those of the EMMD theory
\be
{\cal L}=\sqrt{-g} \Big(R - \ft12 (\partial\phi)^2 - \ft14 e^{a_1\phi} F_{1\2}^2 - \ft14 e^{a_2\phi} F_{2\2}^2\Big)\,,\qquad F_{i\2}=dA_{i\1}\,,\label{emmdlag}
\ee
with equal but opposite dilaton coupling constants $a_2=-a_1$. The generalization to EMMD theory has the advantage of allowing for more general dilaton coupling constants $(a_1,a_2)$, and it also allows us to consider multi-charge configurations in higher dimensions where electromagnetic duality no longer exists. It is also straightforward to generalize this construction to more general EMD-type theories involving multiple dilatonic scalars and Maxwell fields.

\section{Kerr immersed in uniform magnetic field in KK theory}

KK theory is a special EMD theory arising from the KK circle reduction of pure Einstein gravity in five dimensions. The reduction ansatz along the internal $z$ is
\be
d\hat{s}_5^2=e^{-\frac{1}{\sqrt{3}}\phi}ds_4^2+e^{\frac{2}{\sqrt{3}}\phi}
\big(dz+A_{\1}\big)^2\,.\label{KK reduction}
\ee
The resulting four-dimensional theory is an EMD theory with the dilaton coupling constant $a=\sqrt3$.

\subsection{Uniform magnetic electrovacuum}

The Melvin electrovacuum in KK theory can be obtained by performing a suitable rotation mixing the azimuthal coordinate $\varphi$ of $D=4$ Minkwoski spacetime with the fifth (internal) dimension $z$, followed by the KK reduction on $z$. The resulting off-diagonal component gives rise to a magnetic field associated with the four-dimensional KK vector \cite{Dowker:1995gb}. The resulting magnetic field is non-uniform, depending on both the radial and latitudinal coordinates. In this paper, we use the recently constructed $D=4$ Ricci-flat metric as the new seed, namely, the $B$-deformation of the Kerr metric \cite{Ma:2026otg}. For simplicity, we begin with the static vacuum by turning off both the mass and angular momentum parameters. The four-dimensional static Ricci-flat seed solution is \cite{Astorino:2026okd}
\bea
ds_4^2 &=& \fft{1}{(1 + {\cal B} u^2)^2} \Big(- (1 + {\cal B}^2 R^2) dt^2 + \fft{dR^2}{1 + {\cal B} R^2} + \fft{4 du^2}{(1 + {\cal B} u^2)^2} \Big) + 4u^2\, d\varphi^2\,,\nn\\
&=&\frac{\big(1+\sqrt{1+{\cal B}^2r^2(1-x^2)}\big)^2}{4\big(1+{\cal B}^2r^2(1-x^2)\big)^2}
ds^2_{\mathrm{AdS}_3} +\frac{4r^2(1-x^2)}{\big(1+\sqrt{1+{\cal B}^2r^2(1-x^2)}\big)^2}
d\varphi^2\,,\nn\\
%%%
ds^2_{\mathrm{AdS}_3}&=&-(1+{\cal B}^2r^2)dt^2+\frac{dr^2}{1+{\cal B}^2r^2}+r^2\frac{dx^2}{1-x^2}\,.
\label{KKd5B}
\eea
In the above, we present the solutions in two different sets of coordinates. The relation between $(R,u)$ and $(r,x)$ coordinates is
\be
R=\frac{rx}{\sqrt{1+{\cal B}^2r^2(1-x^2)}}\,,\qquad u=\frac{r\sqrt{1-x^2}}{1+\sqrt{1+{\cal B}^2r^2(1-x^2)}}\,.\label{coord RZ}
\ee
It is readily seen from the $(R,u)$ coordinates that this metric is locally equivalent to the Schwarzschild metric in the Kundt class \cite{Astorino:2026okd}. Next, we lift the solution to five dimensions $ds_5^2 = ds_4^2 + dz^2$, which continues to be Ricci flat. We then perform a simple linear coordinate transformation $\phi\rightarrow \phi + { B} z/2$, thereby introducing an off-diagonal term $dz d\varphi$ in the metric. We now perform the KK reduction along the \(z\)-direction using the ansatz \eqref{KK reduction}. We arrive at an exact solution of the KK theory:
\bea
ds_4^2 &=& \fft{\sqrt{H}}{(1 + {\cal B} u^2)^2} \Big(- (1 + {\cal B}^2 R^2) dt^2 + \fft{dR^2}{1 + {\cal B} R^2} + \fft{4 du^2}{(1 + {\cal B} u^2)^2} \Big) + \fft{4u^2}{\sqrt{H}} d\varphi^2\,,\nn\\
&=&\frac{\sqrt{H}}{4}\frac{\big(1+\sqrt{1+{\cal B}^2 r^2 (1-x^2)}\big)^2}{\big(1+{\cal B}^2 r^2 (1-x^2)\big)^2}ds^2_{\mathrm{AdS}_3}
+\frac{4 r^2 (1-x^2)}{\sqrt{H} \big(1+\sqrt{1+{\cal B}^2 r^2 (1-x^2)}\big)^2}d\varphi^2\,,\cr
%%%
A_{\1}&=&\frac{2}{BH}d\varphi\,,\qquad \phi=\frac{\sqrt{3}}{2}\log H\,,\qquad
H=1 + B^2 u^2 = \frac{2 \sqrt{1+B^2 r^2 (1-x^2)}}{1+\sqrt{1+B^2 r^2 (1-x^2)}}\,.\label{BR KK EMD}
\eea
As in the Melvin-like construction starting from Minkowski spacetime, this
new solution also contains a magnetic field but carries no magnetic charge, and therefore represents an electrovacuum solution. Indeed, when ${\cal B}=0$, the solution precisely reduces to the Melvin-like solution in the KK theory \cite{Gibbons:1987ps, Dowker:1993bt,Dowker:1994up,Dowker:1995gb}.

We observe that when ${\cal B}=B$, the resulting electrovacuum has a constant magnetic field, i.e.
\be
F^2=2B^2\,, \qquad \hbox{when}\qquad {\cal B}=B\,.\label{KKFsq}
\ee
This property is analogous to the the BR electrovacuum; however, there is also an important distinction. The BR solution of Einstein-Maxwell theory is locally equivalent to AdS$_2\times S^2$ and consequently has constant Riemann tensor. The metric in KK theory is neither homogeneous nor conformal to a homogeneous spacetime. We find that the Riemann tensor squared is
\be
\mathrm{Riem}^2=\frac{B^4 v^3 (1271 v^2+514 v+71)}{8 (v+1)^5}\,,\qquad v=\sqrt{1+B^2 r^2 (1-x^2)}\,.\label{Riemsq}
\ee
It is now natural to ask whether our KK solution can be extended to general EMD theory, such that the BR solution of Einstein-Maxwell theory is recovered in the $a=0$ limit.
This is one of our goals, and we shall return to this question in the next section.

With this explicit example of the KK solution, we are in a better position to distinguish between an electrovacuum and a fluxbrane. The solution \eqref{BR KK EMD}, written in $(r,x)$ or $(R,u)$ coordinates, is typically viewed as an electrovacuum or a fluxbrane, respectively. To see the difference, we set $B=0={\cal B}$, the metric becomes flat with
\bea
(R,u):\qquad && ds^2 = -dt^2 + dR^2 + 4du^2 + 4u^2 d\varphi^2\,,\nn\\
(r,x):\qquad && ds^2 = -dt^2 + dr^2 + r^2 \Big(\fft{dx^2}{1-x^2} + (1-x^2) d\varphi^2\Big)\,.
\eea
In other words, in the $(R,u)$ coordinates, the flat metric is written in cylindrical coordinates, while in the $(r,x)$ coordinates, it is written in spherical polar coordinates. The distinction becomes physical when we attempt to construct black holes in these backgrounds. A Schwarzschild black hole immersed in an electrovacuum necessarily has comohogeneity two, while a non-extremal fluxbrane can still have cohomogeneity one, typically depending only on coordinate $u$. In this paper, we shall focus on the construction of black holes in electrovacua rather than non-extremal fluxbranes.

\subsection{Kerr in external uniform magnetic field}

The solution-generating technique described the previous subsection can be readily applied starting from the full Ricci-flat $B$-deformed Kerr metric \cite{Ma:2026otg}. For simplicity, we shall assume from the outset that ${\cal B}=B$. We shall not repeat the derivation here and simply present the resulting solution
\bea
ds^2&=&\sqrt{H} \mathbb{L} \Big[-\frac{ P Q (1-x^2)}{P r^4 (1-x^2)-a^2 Q
   x^4}dt^2+\frac{dr^2}{Q}+\frac{1}{P}\frac{dx^2}{1-x^2}\Big]\cr
   &&+\frac{1}{\sqrt{H} \mathbb{L}}\frac{\big[ \big(P r^4
   (1-x^2)-a^2 Q x^4\big)d\varphi+a  \mathbb{M}dt\big]^2}{ \Omega ^4 \big(P r^4
   (1-x^2)-a^2 Q x^4\big)}\,,\cr
   %%%
A_{\1}&=&-\frac{B}{2 H \mathbb{L} \Omega ^4}\Big[ \big(P r^4
   (1-x^2)-a^2 Q x^4\big)d\varphi+a  \mathbb{M}dt\Big]\,,\qquad \phi=\frac{\sqrt{3}}{2}\log H\,,\cr
   %%%
H&=&1 + \frac{B^2}{4 \mathbb{L} \Omega ^4} \Big[P r^4
   (1-x^2)-a^2 Q x^4\Big]\,.\label{kkmagnetickerr}
\eea
Note that here ``$a$'' denotes the rotating parameter of the original Kerr black hole; it should not be confused with the dilaton coupling constant $a$, which is fixed to be $\sqrt3$ in KK theory. The explicit forms of the functions \(\{P,Q,\Omega,I_1, I_2,\mathbb{L},\mathbb{M}\}\) in this solution are given in \cite{Ma:2026otg}, with some of these expressions themselves inherited from \cite{Podolsky:2025tle,Ovcharenko:2025cpm}. For self-completeness, we also present them here:
\bea
\mathbb{M}&=&\frac{1}{2 I_1^3 \Omega }\Bigg\{
\Omega  \Big[I_1 I_2 \mu  r x^2 \big(B^2 \mu  r (B^2 r^2+x^2)+2 I_1 (B^2 r^2+1)\big)-I_1^3 \Sigma  (B^2 r^2 x^2+1)\Big]\cr
&&+(B^2 r^2+1)\Big[
I_1^3 \Sigma  (a^2 B^2 x^2-1)+I_2 \mu  r x^2 \big(I_2 B^4 \mu ^2 r^2 x^2+3 I_1 I_2 B^2 \mu  r x^2\cr
&&-I_1^2 (3 a^2 B^2 x^2+B^2 r^2 x^2-2)\big)
\Big]
\Bigg\}\,,\cr
%%%
\mathbb{L}&=&\frac{1}{4 I_1^2 \Omega ^4}\Big\{
\Sigma  \big[B^4 r^2 \mu ^2 I_2 x^2+I_1^2 \big(2+a^2 B^4 r^2 x^2+B^2 (r^2-a^2 x^2-r^2 x^2)\big)\big]\cr
&&+B^2 r \mu  I_2 x^2 \big(2 a^2 I_1
   x^2+B^2 r^3 \mu  I_2 x^2+2 r^2 I_1 (2-a^2 B^2 x^2)\big)\cr
   &&+2 \Omega  I_1 (\Sigma  I_1+B^2 r^3 \mu  I_2 x^2)
\Big\}\,,\cr
%%%
\Sigma&=&r^2+a^2x^2\,,\qquad
%%%
P=1+B^2\Big(\mu^2\frac{I_2}{I_1^2}-a^2\Big)x^2\,,\qquad Q=(1+B^2r^2)\Delta\,,\cr
%%%
\Omega^2&=&(1+B^2r^2)-B^2\Delta x^2\,,\qquad \Delta=\Big(1-B^2\mu^2\frac{I_2}{I_1^2}\Big)r^2-2\mu\frac{I_2}{I_1}r+a^2\,,\cr
%%%
I_1&=&1-\frac{1}{2}B^2a^2\,,\qquad I_2=1-B^2a^2\,.
\eea
To remove the conical singularity and closed time-like curves from this solution, we perform the following coordinate transformation
\be
\varphi \quad \rightarrow \quad \varphi'=\frac{\varphi}{\sqrt{P_0}}\,,\qquad t\quad\rightarrow \quad t'=t-\frac{2a}{1+\sqrt{I_2}}\frac{\varphi}{\sqrt{P_0}}
\ee
Here we have
\be
P_0=\frac{4 I_2^2 \big[(B^2 \mu ^2+I_1^2)^3 (1+\sqrt{I_2})^{12}+B^2 \mu ^2 I_1^4I_2
   (1-3 \sqrt{I_2})^2 (1-\sqrt{I_2})^6 \big]}{I_1^4
   (1+\sqrt{I_2})^{12} \big[B^2 \mu ^2 (1+\sqrt{I_2})^2+I_1^2 (1+3
   I_2)\big]}\,,
\ee
which ensures that the azimuthal angle $\varphi$ has $2\pi$ period. It is clear that when $\mu=0=a$, the solution reduces to the electrovacuum with a constant magnetic field \eqref{BR KK EMD}. On the other hand, setting $B=0$ simply yields the Kerr metric. Therefore, the exact solution \eqref{kkmagnetickerr} describes a Kerr black hole immersed in an external uniform magnetic field in KK theory.

\subsection{Kerr in conformally-uniform electric field}

As we have discussed earlier, the EMD theory, including KK theory, possesses only the $\mathbb Z_2$ electromagnetic duality \eqref{emduality}. After some rather involved calculations, we find that the electromagnetic dual of \eqref{kkmagnetickerr} is given by
\bea
A_{\1}&=&-\frac{B x}{I_1 r (1+B^2 r^2) \mathbb{L} \Omega ^3}\Bigg\{-a r^2 (1+B^2 r^2) (\mu I_2 r x^2  -I_1 \Sigma )d\varphi\cr
&&+
\Big[
I_1 Q \mathbb{L} \Omega ^2-\frac{a^2 (1+B^2 r^2)(\mu I_2 r
   x^2  -I_1 \Sigma ) (r^2 \mathbb{M}+x^2Q  \mathbb{L} \Omega ^2) }{P r^4 (1-x^2)-a^2 Q x^4}
\Big]dt
\Bigg\}\,,\cr
%%%
\phi&=&-\frac{\sqrt{3}}{2}\log H\,,\label{electricversion}
\eea
while the metric remains unchanged. Setting the mass and rotation parameters $(\mu,a)$ to zero, we obtain the electrovacuum involving only the electric field, given by
\be
A_\1 = -\fft{r x}{2\sqrt{1+B^2 r^2 (1-x^2)}} dt\,,\qquad \phi=
-\fft{\sqrt3}2 \log\fft{2\sqrt{1 + B^2 r^2 (1-x^2)}}{1 + \sqrt{1 + B^2 r^2 (1-x^2)}}\,,
\ee
while the metric is the same as that given in \eqref{BR KK EMD}. The electric field is not uniform, but is conformally uniform as can be readily verified by
\be
e^{2\sqrt3 \phi} F_\2^2\Big|_{\rm EF} = F_\2^2\Big|_{\rm PF}=-2B^2\,.\label{EFPFKK}
\ee
(See section \ref{sec:emdreview} for details.)

\subsection{Uplifting to new $D=5$ Ricci-flat metric}

It is clear that lifting the solution \eqref{kkmagnetickerr} back to $D=5$ is trivial, simply yielding a direct product of Kerr metric and an $S^1$. However, lifting the electric version \eqref{electricversion} back to $D=5$ gives rise to a new Ricci-flat metric, given by
\bea
ds^2&=&H \mathbb{L} \Big[-\frac{ P Q (1-x^2)}{P r^4 (1-x^2)-a^2 Q
   x^4}dt^2+\frac{dr^2}{Q}+\frac{1}{P}\frac{dx^2}{1-x^2}\Big]\cr
   &&+\frac{\big[ \big(P r^4
   (1-x^2)-a^2 Q x^4\big)d\varphi+a  \mathbb{M}dt\big]^2}{ \mathbb{L} \Omega ^4 \big(P r^4
   (1-x^2)-a^2 Q x^4\big)} + \frac{1}{H} (dz + A_\1)^2\,,
\eea
where $A_\1$ is given by \eqref{electricversion}. It is straightforward to verify that this metric is Ricci flat. It is worth commenting that if we again make the same linear coordinate between $z$ and $\varphi$, and then perform the KK reduction along $z$, we will arrive at a Kerr black hole in both external electric and magnetic fields in KK theory. However, such construction will yield an unavoidable swirling term in the metric such that the solution is stationary even when we turn off the Kerr rotating parameter $a$. Furthermore the dilaton cannot be decoupled to give rise to a solution of Einstein-Maxwell theory. We consider such configuration undesirable and we therefore shall not present the results here.

\section{Schwarzschild in uniform magenetic fields in EMD theories}

Having constructed the Kerr black hole in either an external uniform magnetic field or a conformally-uniform electric field in KK theory, a natural next step is to construct analogous solutions in EMD theories with a generic dilaton coupling constant $a$. (Not to be confused with rotating parameter $a$ discussed earlier.) Since solution-generating techniques are no longer applicable and exact solutions of charged rotating black holes in EMD theories are generally unknown, we shall consider only the static case. Following the experience we gained from the KK theory discussed in the previous section and from charged black holes in EMD theories, we find, after some guesswork, that solutions describing a Schwarzschild black hole immersed in a uniform magnetic field are reasonably simple in EMD theories, and are given by
\bea
ds^2&=&H^{\frac{2}{a^2+1}}\frac{(1+\Omega+B^2 \mu  r x^2)^2}{4\Omega^4}\Big[-\frac{Q}{r^2}dt^2+\frac{r^2}{Q}dr^2
+\frac{r^2}{P}\frac{dx^2}{1-x^2}\Big]\cr
&&+H^{-\frac{2}{a^2+1}}\frac{4r^2P(1-x^2)}{(1+\Omega+B^2 \mu  r x^2)^2}d\varphi^2\,,\cr
%%%
A_{\1}&=&\frac{4}{\sqrt{a^2+1}BH}d\varphi\,,\qquad \phi=\frac{2a}{a^2+1}\log H\,,\qquad H=\frac{2\Omega}{1+\Omega+B^2 \mu  r x^2}\,,\cr
%%%
P&=&1+B^2 \mu ^2 x^2\,,\qquad \Omega^2=1+B^2 \Big(\mu  r  \big(B^2 \mu  r+2\big)x^2+r^2 \big(1-x^2\big)\Big)\,,\cr
%%%
Q&=&r^2\big(1+B^2 r^2\big) \Big(1-B^2 \mu ^2-\frac{2 \mu }{r}\Big)\,.\label{emdbh}
\eea
When $B=0$, the solution describes the Schwarzschild black hole. For $\mu=0$, it describes an electrovacuum with a uniform magnetic field. The squared Riemann tensor and Maxwell field strength for the electrovacuum ($\mu=0$) are given by
\bea
\mathrm{Riem}^2&=&\frac{B^4 u^{\frac{4 a^2}{a^2+1}}}{(a^2+1)^4} \Big(\frac{2}{u+1}\Big)^{\frac{2 (3 a^2+1)}{a^2+1}}\Big[(12 a^8+8 a^6+7 a^4+6 a^2+2) u^2\cr
&&+2 (8 a^6+3 a^4+4 a^2+2) u+7 a^4+2 a^2+2\Big]\,,\qquad F^2=\frac{8B^2}{a^2+1}\,,
\eea
where $u$ is given in \eqref{Riemsq}. Interestingly the leading falloff of $u$ at large $u$ is $1/u^2$, independent of the dilaton coupling constant $a$.

When the dilaton coupling constant $a=\sqrt3$, the solution reduces to the corresponding static black hole in KK theory discussed earlier. For \(a=0\), the dilaton decouples, and the solution reduces to the static case of the KBR black hole obtained in \cite{Podolsky:2025tle, Ovcharenko:2025cpm}. We thus complete the main goal of this paper, namely, to unify Schwarzschild black holes immersed in external uniform magnetic fields in both Einstein-Maxwell and KK theories within the broader framework of general EMD theories. It is important to note that, in the black hole construction, the concepts of the electrovacuum and fluxbrane become distinct. The above black hole solution would be very unnatural in the fluxbrane $(R, u)$ coordinates.

The external electric field configuration can be obtained through the $\mathbb Z_2$ electromagnetic duality. The metric remains unchanged, while the Maxwell field and dilaton are transformed into
\be
A_{\1}=\frac{2BQx}{\sqrt{a^2+1}r(1+B^2r^2)\Omega}dt\,,\qquad \phi=-\frac{2a}{a^2+1}\log H\,.
\ee
If we turn off the mass parameter $\mu$, the electrovacuum involves a conformally-uniform electric field, given by
\be
e^{2a \phi} F_\2^2\Big|_{\rm EF} = F_\2^2\Big|_{\rm PF}=-\fft{8B^2}{a^2 +1}\,.\label{EFPFEMD}
\ee

\section{Electromagnetic electrovacua in EMD theories}

In the previous sections, we constructed directly exact solutions in EMD theories, describing the Schwarzschild black hole immersed in either a uniform external magnetic field or a conformally-uniform electric field. In this section, we construct electrovacua in EMD theories that contain both electric and magnetic fields with independent parameters. Such construction is trivial in Einstein-Maxwell theory owing to the continuous $SO(2)$ electromagnetic duality, but highly nontrivial in the present case, where we have only the discrete $\mathbb Z_2$ symmetry. We therefore construct the solutions directly from a suitable ansatz, rather than using a solution-generating technique, which, to our knowledge, is not available for the general EMD theories, when both electric and magnetic fields are involved.

\subsection{Electrovacua in fluxbrane coordinates}

As we have seen earlier, electrovacua and fluxbranes are locally equivalent and the solutions of the latter appear significantly simpler. We therefore write the metric \eqref{emdbh} with the vanishing blackening parameter ($\mu=0$) in the simpler $(R,u)$ coordinates:
\be
ds^2=\frac{H^{\frac{2}{1+a^2}}}{(1+B^2 u^2)^2} \Big[- (1+B^2R^2)dt^2+\frac{dR^2}{ 1+B^2R^2}+\frac{4 du^2}{(1+B^2 u^2)^2}\Big]+4 H^{-\frac{2}{1+a^2}} u^2 d\varphi^2\,.
\ee
Here \(H=1+B^2u^2\) is a univariate polynomial in \(u\), making the solution easier to generalize. In other words, although the black hole solution with $\mu\ne 0$ is of cohomogeneity two, the electrovacuum solution with $\mu=0$ becomes cohomogeneity one, depending only on the coordinate $u$. The Maxwell field and dilaton in the magnetic and electric field configurations are respectively given by
\bea
\mathrm{Magnetic}&:&\qquad A_{\1}=\frac{4}{\sqrt{a^2+1}BH}d\varphi\,,\qquad \phi=\frac{2a}{a^2+1}\log H\,,\cr
%%%
\mathrm{Electric}&:&\qquad A_{\1}=-\frac{2BR}{\sqrt{a^2+1}}dt\,,\qquad \phi=-\frac{2a}{a^2+1}\log H\,.
\eea
Furthermore, we note that the parameter $B$ appearing in the function $H$ can differ from the $B$ appearing in the metric, which we denote by ${\cal B}$, as in the case of \eqref{BR KK EMD}.

\subsection{General ansatz and consistent equations}

We are now in the position to construct electrovacua involving both electric and magnetic fields. To achieve this, we need to consider an ansatz with three independent parameters, associated with the metric, electric field and magnetic field. We therefore consider the following cohomogeneity-one ansatz
\bea
ds^2&=&\frac{(H_1H_2)^{\frac{2}{1+a^2}}}{(1+\mathcal{B}^2 u^2)^2} \Big[- (1+\mathcal{B}^2R^2)dt^2+\frac{dR^2}{ 1+\mathcal{B}^2R^2}+\frac{4 du^2}{(1+\mathcal{B}^2 u^2)^2}\Big]+4 (H_1 H_2)^{-\frac{2}{1+a^2}}  u^2 d\varphi^2\,,\cr
%%%
F_{\2}&=&\frac{2E}{\sqrt{a^2+1}}dt\wedge dR+\frac{8 B H_2^{\frac{2 (a^2-1)}{a^2+1}}}{\sqrt{a^2+1} H_1^2}\,u du\wedge d\varphi\,,\qquad \phi=\frac{2a}{a^2+1}\log\frac{H_1}{H_2}\,.
\eea
Here \(H_{1,2}\) remain univariate functions of \(u\). Note that we have introduced a new parameter \(\mathcal{B}\) in addition to the electromagnetic field parameters \(B\) and \(E\). For this ansatz, the equations of motion reduce to three ordinary differential equations, two of which are second-order differential equations, while the third is the consistent first-order Hamiltonian constraint. They are given by
\bea
&&\!\!\!\!\!\!\!4 B^2\frac{ H_2^{\frac{2 (a^2-1)}{a^2+1}}}{H_1^2}-\frac{H_1'}{u H_1}+\frac{H_1'^2}{H_1^2}-\frac{H_1''}{H_1}=0\,,\qquad 4 E^2\frac{ H_1^{\frac{2 (a^2-1)}{a^2+1}}}{H_2^2}-\frac{H_2'}{u H_2}+\frac{H_2'^2}{H_2^2}-\frac{H_2''}{H_2}=0\,,\cr
%%%
&&\!\!\!\!\!\!\mathcal{H}\equiv \frac{ 4E^2 H_1^{\frac{2 (a^2-1)}{a^2+1}}}{H_2^2}+\frac{4B^2 H_2^{\frac{2 (a^2-1)}{a^2+1}}}{H_1^2}-\frac{2 H_1'}{u
   H_1}-\frac{2 H_2'}{u H_2}-\frac{2 (a^2-1) H_1' H_2'}{(a^2+1) H_1
   H_2}+\frac{H_1'^2}{H_1^2}+\frac{H_2'^2}{H_2^2}=0\,,\label{EMD vacuum}
\eea
where a prime denotes a derivative with respect to coordinate $u$. It is easy to verify that ${\cal H}' = -2{\cal H}/u$ upon substituting the second-order equation into ${\cal H}'$. Thus, solutions to solve the system of of equations \eqref{EMD vacuum} must exist.

\subsection{Special analytical solutions}

For \(a=1\) and \(a=\sqrt{3}\), the above equations admit analytical solutions for general $EB\ne0$, which are given by
\bea
a=1&:&\qquad H_1=1+B^2u^2\,,\qquad H_2=1+E^2u^2\,,\cr
%%%
a=\sqrt{3}&:&\qquad H_1=1+B^2u^2+\frac{1}{4}B^2E^2u^4\,,\qquad H_2=1+E^2u^2+\frac{1}{4}B^2E^2u^4\,.
\eea
For general dilaton coupling constants with generic $(E,B)$, analytic solutions are unaccessible, and one must resort to numerical methods, which we shall not discuss in this paper. For completeness, we also express the solutions in the electrovacuum  \(\{r,x\}\) coordinates by using the coordinate transformation \eqref{coord RZ}. We have
\bea
ds^2&=&\frac{H_1^{\frac{2}{a^2+1}}H_2^{\frac{2}{a^2+1}}}{4}\frac{\big(1+\sqrt{1+\mathcal{B}^2 r^2 (1-x^2)}\big)^2}{\big(1+\mathcal{B}^2 r^2 (1-x^2)\big)^2}ds^2_{\mathrm{AdS}_3}\cr
&&+H_1^{-\frac{2}{a^2+1}}H_2^{-\frac{2}{a^2+1}}\frac{4 r^2 (1-x^2)}{ \big(1+\sqrt{1+\mathcal{B}^2 r^2 (1-x^2)}\big)^2}d\varphi^2\,,\cr
%%%
ds^2_{\mathrm{AdS}_3}&=&-(1+\mathcal{B}^2r^2)dt^2+\frac{dr^2}{1+\mathcal{B}^2r^2}+r^2\frac{dx^2}{1-x^2}\,,\qquad \phi=\frac{2a}{a^2+1}\log\frac{H_1}{H_2}\,.
\eea
For the two exact solutions obtained above, we can explicitly express \(A_{\1}\), rather than its field strength, as follows
\bea
a=1&:&\quad A_{\1}=-\frac{\sqrt{2}Erx}{\sqrt{1+\mathcal{B}^2 r^2 (1-x^2)}}dt-\frac{2\sqrt{2}}{BH_{1,a=1}}d\varphi\,,\\
%%%
a=\sqrt{3}&:& \quad A_{\1}=-\frac{Erx}{\sqrt{1+\mathcal{B}^2 r^2 (1-x^2)}}dt-\frac{4}{BH_{1,a=\sqrt{3}}}\Big[
1+\frac{B^2r^2(1-x^2)}{2\big(1+\sqrt{1+\mathcal{B}^2 r^2 (1-x^2)}\big)^2}
\Big]d\varphi\,.\nn
\eea
It is worth emphasizing that \(\mathcal{B}\) is an independent parameter here, in which case neither $F_\2^2$ or $F_\2^2|_{\rm PF}$ is constant. When \(\mathcal{B}=B\), the invariant \(F_\2^2\) becomes a constant when $E=0$; when \(\mathcal{B}=E\), the quantity
\(F_\2^2|_{\rm PF}\) is constant when $B=0$, recovering either the purely magnetic or electric cases constructed earlier.

\subsection{Einstein-Maxwell limit}

A special class of the electrovacua discussed above corresponds to those in Einstein-Maxwell gravity. These solutions can be obtained from taking $a=0$, in which case, the dilaton scalar decouples from our ansatz. Alternatively, we may take \(E=B\rightarrow B/\sqrt{1+a^2}\), in which case, we find an exact solution with arbitrary \(a\), given by
\be
A_{\1}=-2BRdt-\frac{2H_1^{-\frac{2}{a^2+1}}}{B}d\varphi\,,\qquad
%%%%
H_1=H_2=\Big(1+2B u^2  \Big)^{\frac{1}{2} (a^2+1)}.
\ee
We can see that, in this case, the dilaton also completely decouples, and the above solution reduces to a solution of the Einstein-Maxwell theory. To write the solution in a standard form, we introduce a function \(H\),
\be
H=1+2B u^2=1+\frac{2 B^2 r^2 (1-x^2)  }{\big(1+\sqrt{1+\mathcal{B}^2 r^2 (1-x^2)}\big)^2}\,.
\ee
The electrovacuum solution in Einstein-Maxwell gravity then becomes
\bea
ds^2&=&H^2\frac{\big(1+\sqrt{1+\mathcal{B}^2 r^2 (1-x^2)}\big)^2}{4 \big(1+\mathcal{B}^2 r^2 (1-x^2)\big)^2}\Big[
-(1+\mathcal{B}^2 r^2) dt^2+\frac{dr^2}{1+\mathcal{B}^2 r^2}+r^2\frac{ dx^2}{1-x^2}
\Big]\cr
&&+\frac{4 r^2 (1-x^2)}{H^2 \big(1+\sqrt{1+\mathcal{B}^2 r^2 (1-x^2)}\big)^2}d\varphi^2\,,\cr
%%%
A_{\1}&=&-\frac{2 B r x}{\sqrt{1+\mathcal{B}^2 r^2 (1-x^2)}}dt-\frac{2}{B H}d\varphi\,.
\eea
When $B={\cal B}$, this reduces to the standard the BR electrovacuum where both electric and magnetic fields contribute equally.

Before finishing this subsection, we would like to comment that we obtain our electrovacuum solutions by direct construction rather than making use a solution generating technique, even in KK theory. At first sight, for the KK theory, one might expect to use the solution generating technique first to obtain the electrovacuum with an external magnetic field, as we did earlier.  We can perform the $\mathbb Z_2$ electromagnetic duality to obtain an electrovacuum with an electric field. We can lift the solution back to $D=5$ and then perform the same procedure to obtain the KK solution with independent electric and magnetic fields. However, it turns out that such construction inevitably generates an off-diagonal ($dt d\varphi$) term in $D=4$, rendering the metric stationary rather than static as in our solutions. Such stationary configurations are referred to as swirling \cite{Taub:1950ez}. (See our earlier comments at the end of Section 3.) Furthermore, we find that the dilaton cannot be decoupled in this process even when the electric and magnetic field parameters $E$ and $B$ are set to equal. It would be of great interest to investigate whether there exist other forms of solution-generating techniques that can give rise not only to our electrovacua, but also to black holes as well, at least in the KK theory. In the next subsection, we shall instead directly construct such a solution for special values of $a$.

\subsection{Schwarzschild black hole in external electromagnetic fields}

We have so far constructed electrovacua in EMD theories that involve independent external electric and magnetic fields. However, we have been unable to introduce the blackening parameter $\mu$ and construct the Schwarzschild black hole immersed in such electrovacua for general dilaton couplings, even at the level of ansatz.  For two special cases, namely $a=1$ and $a=\sqrt3$, both of which are related to string theories, we find that not only do analytic electrovacua exist, but static black hole solutions also exist. For $a=1$, we find that the static black hole solution is
\bea
ds^2&=&H_1 H_2\frac{(1+\Omega+{\cal B}^2 \mu  r x^2)^2}{4\Omega^4}\Big[-\frac{Q}{r^2}dt^2+\frac{r^2}{Q}dr^2
+\frac{r^2}{P}\frac{dx^2}{1-x^2}\Big]\cr
&&+\frac{4r^2P(1-x^2)}{H_1 H_2 (1+\Omega+{\cal B}^2 \mu  r x^2)^2}d\varphi^2\,,\cr
%%%
A_{\1}&=&-\frac{\sqrt2 E\,Q\,x}{r(1 + {\cal B}^2 r^2)\Omega} dt-\frac{2\sqrt2}{BH_1}d\varphi\,,\qquad \phi=\log \frac{H_1}{H_2}\,,\cr
%%%
P&=&1+{\cal B}^2 \mu ^2 x^2\,,\qquad \Omega^2=1+{\cal B}^2 \Big(\mu  r  \big({\cal B}^2 \mu  r+2\big)x^2+r^2 \big(1-x^2\big)\Big)\,,\cr
%%%
Q&=&r^2\big(1+{\cal B}^2 r^2\big) \Big(1-{\cal B}^2 \mu ^2-\frac{2 \mu }{r}\Big)\,,\label{emda=1bh}
\eea
where the functions $H_1$ and $H_2$ are given by
\be
H_1 = 1 + \frac{B^2 r^2(1-x^2) P}{(1 + {\cal B}^2 \mu r x^2 + \Omega)^2}\,,\qquad
H_2 = 1 + \frac{E^2 r^2(1-x^2) P}{(1 + {\cal B}^2 \mu r x^2 + \Omega)^2}\,.
\ee
For $a=\sqrt3$, the black hole solution is more complicated, and is given by
\bea
ds^2&=&\sqrt{H_1 H_2}\frac{(1+\Omega+{\cal B}^2 \mu  r x^2)^2}{4\Omega^4}\Big[-\frac{Q}{r^2}dt^2+\frac{r^2}{Q}dr^2
+\frac{r^2}{P}\frac{dx^2}{1-x^2}\Big]\cr
&&+\frac{4r^2P(1-x^2)}{\sqrt{H_1 H_2} (1+\Omega+{\cal B}^2 \mu  r x^2)^2}d\varphi^2\,,\cr
%%%
A_{\1}&=& - \frac{E Q x}{r(1 + {\cal B}^2 r^2)\Omega} dt +
\fft{4}{B H_1} \Big(1 + \frac{B^2 r^2 (1-x^2)P}{2(1 + {\cal B}^2 \mu r x^2 + \Omega)^2}\Big) d\varphi\,,\quad \phi=\frac{\sqrt3}{2}\log \frac{H_1}{H_2}\,,
\eea
where $(P,Q,\Omega)$ is the same as in the $a=1$ case. The functions $(H_1,H_2)$ are now given by
\bea
H_1 &=& 1 + B^2 \frac{r^2 (1-x^2) P}{(1 + {\cal B}^2 \mu r x^2 + \Omega)^2} +
\frac{B^2 E^2}{4} \Big(\frac{r^2 (1-x^2) P}{(1 + {\cal B}^2 \mu r x^2 + \Omega)^2}\Big)^2\,,\cr
H_2 &=& 1 + E^2 \frac{r^2 (1-x^2) P}{(1 + {\cal B}^2 \mu r x^2 + \Omega)^2} +
\frac{B^2 E^2}{4} \Big(\frac{r^2 (1-x^2) P}{(1 + {\cal B}^2 \mu r x^2 + \Omega)^2}\Big)^2\,.
\eea
Note that the above successful black hole constructions were obtained purely through intuitive guesswork without a systematic understanding. Consequently, we are unable to generalize the construction to arbitrary dilaton coupling constants beyond $a=1$ and $a=\sqrt3$. It is worth noting that when $E=B ={\cal B}$, both $a=1$ and $\sqrt3$ solutions reduce to the same static KBR solution with the decoupled dilaton.

\section{Electrovacua and black holes in EMMD theories}

\subsection{Electrovacua or fluxbranes}

The electrovacua involving both electric and magnetic fields in EMD theories can be viewed as purely electric or magnetic fields in EMMD theories with two equal but opposite dilaton coupling constants. This allows us to construct electrovacua in the more general theories involving two Maxwell fields with two independent coupling constant $(a_1,a_2)$. The Lagrangian of EMMD theory is given by \eqref{emmdlag}. Restricting to the purely electric case, the corresponding ansatz in the fluxbrane $\{R,u\}$ coordinates, associated with fluxbranes, is given by
\bea
&&\!\!\!ds^2=\frac{H_1^{\frac{2}{a_1^2+1}}H_2^{\frac{2}{a_2^2+1}}}{(1+\mathcal{B}^2 u^2)^2} \Big[- (1+\mathcal{B}^2R^2)dt^2+\frac{dR^2}{ 1+\mathcal{B}^2R^2}+\frac{4 du^2}{(1+\mathcal{B}^2 u^2)^2}\Big]+4u^2 H_1^{-\frac{2}{a_1^2+1}}H_2^{-\frac{2}{a_2^2+1}}  d\varphi^2\,,\cr
%%%
&&\!\!\!F_{1,2\2}=\frac{2 E_{1,2}}{\sqrt{a_{1,2}^2+1}}dt\wedge dR\,,\qquad \phi=-\frac{2a_1}{a_1^2+1}\log H_1-\frac{2a_2}{a_2^2+1}\log H_2\,.
\eea
Similar to the EMD case \eqref{EMD vacuum}, we also obtain two second-order ordinary differential equations
\be
0=\frac{4 E_1^2}{H_1^2 H_2^{\frac{2 (a_1 a_2+1)}{a_2^2+1}}}-\frac{H_1'}{u H_1}+\frac{H_1'^2}{H_1^2}-\frac{H_1''}{H_1}\,,\qquad
   %%%
0=\frac{4 E_2^2}{H_1^{\frac{2 (a_1 a_2+1)}{a_1^2+1}} H_2^2}-\frac{H_2'}{u H_2}+\frac{H_2'^2}{H_2^2}-\frac{H_2''}{H_2}\,,\label{secondordereom}
\ee
together with the first-order Hamiltonian constraint
\bea
\mathcal{H} &\equiv& \frac{4 E_1^2}{(a_1^2+1) H_1^2 H_2^{\frac{2 (a_1 a_2+1)}{a_2^2+1}}}+\frac{4 E_2^2}{(a_2^2+1) H_1^{\frac{2
   (a_1 a_2+1)}{a_1^2+1}} H_2^2}-\frac{2 H_1'}{u (a_1^2+1) H_1}-\frac{2 H_2'}{u (a_2^2+1) H_2}\cr
   &&+\frac{2
   (a_1 a_2+1) H_1' H_2'}{(a_1^2+1) (a_2^2+1) H_1 H_2}+\frac{H_1'^2}{(a_1^2+1)
   H_1^2}+\frac{H_2'^2}{(a_2^2+1) H_2^2}=0\,.\label{firstordercons}
\eea
This Hamiltonian constraint is consistent with the two second-order differential equations, as can be easily verified by ${\cal H}'=-2{\cal H}/u$. It is interesting to note that all three differential equations are independent of the parameter ${\cal B}$.

The second-order equations in \eqref{secondordereom} can be recast as Toda-like equations. To see this, we first introduce functions $(q_1,q_2)$, defined by
\be
H_1 = \lambda_1 e^{-q_1 - \fft{(1+a_1^2) a_2 \log u}{a_1-a_2}}\,,\qquad
H_2 = \lambda_2 e^{-q_2 - \fft{(1+a_2^2) a_1 \log u}{a_2-a_1}}\,,
\ee
where $(\lambda_1,\lambda_2)$ are given by
\be
E_1^2 = \fft14 \lambda_1^2 \lambda_2^{\frac{2(1+a_1 a_2)}{1+a_2^2}}\,,\qquad
E_2^2 = \fft14 \lambda_2^2 \lambda_1^{\frac{2(1+a_1 a_2)}{1+a_1^2}}\,.
\ee
We then redefine the coordinate $u$ as $\eta= \log u$, in terms of which the second-order differential equations become one-dimensional Toda-like equations
\be
\ddot q_1 = - e^{2 q_1 + \fft{2(1+a_1 a_2)}{1+a_2^2} q_2}\,,\qquad
\ddot q_2 = - e^{2 q_2 + \fft{2(1+a_1 a_2)}{1+a_1^2} q_1}\,.\label{todalike}
\ee
Here a dot denotes a derivative with respect to coordinate $\eta$. The Hamiltonian constraint becomes
\be
\fft{\dot q_1^2-\ddot q_1}{1+a_1^2} + \fft{\dot q_2^2-\ddot q_2}{1+a_2^2} +
\fft{2(1+a_1 a_2)}{(1+a_1^2)(1+a_2^2)} \dot q_1 \dot q_2=1\,.
\ee
For suitable choices of dilaton couplings $(a_1,a_2)$, the above equations become Toda equations associated with rank-2 Lie groups, namely $A_2$, $D_2$, $B_2$ and $G_2$. Specifically, when \(a_1=-a_2=1\) or \(\sqrt{3}\), the equations correspond to $D_2$ and $A_2$ Toda equations, respectively. The exact solutions are
\bea
a_1=-a_2=1&:&\qquad H_1=1+E_1^2u^2\,,\qquad H_2=1+E_2^2u^2\,,\cr
%%%
a_1=-a_2=\sqrt{3}&:&\qquad H_1=1+E_1^2u^2+\frac{1}{4}E_1^2E_2^2u^4\,,\qquad H_2=1+E_2^2u^2+\frac{1}{4}E_1^2E_2^2u^4\,.
\eea
For $a_1 = 3$ and $a_2 = -2$, the field equations corresponds to the $B_2$ Toda system, with the solution
\bea
H_1&=&1+E_1^2 u^2+\frac{1}{2} E_1^2 E_2^2 u^4+\frac{1}{9} E_1^2 E_2^4 u^6+\frac{1}{144} E_1^4 E_2^4 u^8\,,\cr
%%%
H_2&=&1+E_2^2 u^2+\frac{1}{4} E_1^2 E_2^2 u^4+\frac{1}{36} E_1^2 E_2^4 u^6\,.
\eea
Finally, for the parameter choice $a_1 = 3\sqrt{3}$ and $a_2 = -\frac{5}{\sqrt{3}}$, the field equations take the form of the $G_2$ Toda system, which admits the following exact solution
\bea
H_1&=&1+E_1^2 u^2+\frac{3}{4} E_1^2 E_2^2 u^4+\frac{1}{3} E_1^2 E_2^4 u^6+\frac{7}{600} E_1^4 E_2^6 u^{10}+\frac{E_1^6 E_2^8
   }{10800}u^{14}+\frac{E_1^6 E_2^{10} }{172800}u^{16}\cr
   &&+\frac{E_1^6 E_2^{12} }{4665600}u^{18}+\frac{E_1^8
   E_2^{12}}{466560000}u^{20}+\frac{E_1^2 E_2^4}{48} (E_1^2+3 E_2^2) u^8+\frac{E_1^4 E_2^6 (27 E_1^2+25
   E_2^2) }{43200}u^{12}\,,\cr
   %%%
H_2&=&1+E_2^2 u^2+\frac{1}{4} E_1^2 E_2^2 u^4+\frac{1}{18} E_1^2 E_2^4 u^6+\frac{1}{144} E_1^2 E_2^6 u^8+\frac{E_1^4 E_2^6
   }{3600}u^{10}+\frac{E_1^4 E_2^8 }{129600}u^{12}\,.
\eea
For more general $(a_1,a_2)$ cases with general $(E_1,E_2)$ parameters, exact solutions are not accessible. It is worth commenting that Toda equations arise naturally in string theory
\cite{Lu:1997rd,Lu:2013uia}. Classes of Toda fluxbranes and dilatonic charged black holes were given in \cite{Ivashchuk:2013jja}. However, our fluxbrane ansatz differs from those in \cite{Ivashchuk:2013jja}, and yet the equations of motion are nevertheless governed by the same Toda-like system. It would be of great interest to investigate the connection between these constructions.

\subsection{Schwarzschild black holes in the electrovacua}

Based on our experience in the EMD theory, where exact solutions for the Schwarzschild black hole immersed in both electric and magnetic fields were constructed for $a=1$ and $a=\sqrt3$, we expect that such black hole solutions should also exist in the analytic electrovacua associated with the Toda equations of rank-2 Lie groups. Again, we apply our guesswork and propose the follow ansatz of black holes
\bea
ds^2&=&H_1^{\frac{2}{a_1^2+1}}H_2^{\frac{2}{a_2^2+1}}\frac{(1+\Omega+{\cal B}^2 \mu  r x^2)^2}{4\Omega^4}\Big[-\frac{Q}{r^2}dt^2+\frac{r^2}{Q}dr^2
+\frac{r^2}{P}\frac{dx^2}{1-x^2}\Big]\cr
&&+H_1^{-\frac{2}{a_1^2+1}}H_2^{-\frac{2}{a_2^2+1}} \frac{4r^2P(1-x^2)}{(1+\Omega+{\cal B}^2 \mu  r x^2)^2}d\varphi^2\,,\cr
%%%
A_{\1,i}&=& - \frac{2E_i Q x}{\sqrt{a_i^2+1}\, r(1 + {\cal B}^2 r^2)\Omega} dt\,,\qquad \phi=-\frac{2a_1}{a_1^2+1}\log H_1-\frac{2a_2}{a_2^2+1}\log H_2\,,
\eea
where $(P,Q,\Omega)$ are the same as those given in \eqref{emda=1bh}. The functions $(H_1,H_2)$ are given by the above corresponding $H_i(u)$ above, with $u$ replaced by
\be
u^2\rightarrow \frac{r^2 (1-x^2) P}{(1 + {\cal B}^2 \mu r x^2 + \Omega)^2}\,.
\ee
It then becomes a straightforward exercise to verify that the above ansatz indeed satisfies all the equations of motion for the special $(a_1,a_2)$ values discussed above. We therefore have constructed the Schwarzschild black holes immersed in two external independent electric fields governed by Toda equations associated with all rank-2 Lie groups. These cohomogeneity-two solutions are completely different from the cohomogeneity-one dilatonic charged black holes assocaited Toda equations of rank-2 Lie groups constructed in \cite{Ivashchuk:2013jja,Lu:2026lkv}.

\section{Conclusions}

In this paper, we considered the recently obtained \(B\)-deformed Ricci-flat solution \cite{Ma:2026ima}, and applied the standard solution-generating technique to obtain the Kerr black hole immersed in an external uniform magnetic field in KK theory. Using this as our starting point, we abandoned the solution-generating technique and instead construct new exact solutions directly, describing the Schwarzschild black hole immersed in a uniform magnetic field. The $SO(2)$ electromagnetic duality of Einstein-Maxwell gravity breaks down to a $\mathbb Z_2$ symmetry in EMD theories, and the electric dual is conformally uniform as we explained in Section \ref{sec:emdreview}.

We then constructed electrovacua with independent external electric and magnetic fields in EMD theories, and generalize the analogous construction to the more general class of EMMD theories. We find that the governing equations of motion for the two-parameter electrovacua can be reduced to Toda-like equations, which become Toda equations for suitable choices of dilaton coupling constants. We therefore obtained analytic solutions for electrovacua associated with all the rank-2 Lie groups. Furthermore, we constructed exact solutions of the Schwarzschild black hole immersed in these electrovacua.

In this paper, we focused on the construction of local solutions without analysing their global structure. One reason is that we have constructed a large number of black hole solutions, many of which are beyond what the title of this paper suggests. They
require further investigation in a separate study. On the other hand, for the general case involving both electric and magnetic fields, exact solutions exist only for some very special values of dilaton coupling constants, and therefore we are still lacking a complete picture. The string-inspired EMD gravities and their generalizations such as EMMD-type theories are very special, and it was recently proposed that black hole thermodynamics can be derived without having to construct back hole solutions at all \cite{Lu:2025eub,Yang:2025rud}. The conclusion was verified for a large number of black holes and $p$-branes, both numerical or analytic \cite{Han:2026bim,Lu:2026lkv}. It would be of great interest to investigate whether analogous approach can applied to black holes immersed in elecrovacua in EMD-like theories as well.

\section*{Acknowledgement}

L.M.~is supported in part by National Natural Science Foundation of China (NSFC) grant No.~12447138, Postdoctoral Fellowship Program of CPSF Grant No.~GZC20241211, the China Postdoctoral Science Foundation under Grant No.~2024M762338 and the National Key Research and Development Program No.~2022YFE0134300. P.Y.W.~and H.L.~are supported in part by the NSFC grants No.~12375052 and No.~11935009. The work is also supported in part by the Tianjin University Self-Innovation Fund Extreme Basic Research Project Grant No.~2025XJ21-0007.

\end{document}